\PassOptionsToPackage{unicode}{hyperref}
\PassOptionsToPackage{hyphens}{url}
\PassOptionsToPackage{dvipsnames,svgnames,x11names}{xcolor}
\documentclass[
]{article}
\usepackage{amsmath,amssymb}
\usepackage{iftex}
\ifPDFTeX
  \usepackage[T1]{fontenc}
  \usepackage[utf8]{inputenc}
  \usepackage{textcomp} % provide euro and other symbols
\else % if luatex or xetex
  \usepackage{unicode-math} % this also loads fontspec
  \defaultfontfeatures{Scale=MatchLowercase}
  \defaultfontfeatures[\rmfamily]{Ligatures=TeX,Scale=1}
\fi
\usepackage{lmodern}
\ifPDFTeX\else
\fi
\IfFileExists{upquote.sty}{\usepackage{upquote}}{}
\IfFileExists{microtype.sty}{% use microtype if available
  \usepackage[]{microtype}
  \UseMicrotypeSet[protrusion]{basicmath} % disable protrusion for tt fonts
}{}
\makeatletter
\@ifundefined{KOMAClassName}{% if non-KOMA class
  \IfFileExists{parskip.sty}{%
    \usepackage{parskip}
  }{% else
    \setlength{\parindent}{0pt}
    \setlength{\parskip}{6pt plus 2pt minus 1pt}}
}{% if KOMA class
  \KOMAoptions{parskip=half}}
\makeatother
\usepackage{xcolor}
\usepackage[margin=3cm]{geometry}
\usepackage{graphicx}
\makeatletter
\newsavebox\pandoc@box
\newcommand*\pandocbounded[1]{% scales image to fit in text height/width
  \sbox\pandoc@box{#1}%
  \Gscale@div\@tempa{\textheight}{\dimexpr\ht\pandoc@box+\dp\pandoc@box\relax}%
  \Gscale@div\@tempb{\linewidth}{\wd\pandoc@box}%
  \ifdim\@tempb\p@<\@tempa\p@\let\@tempa\@tempb\fi% select the smaller of both
  \ifdim\@tempa\p@<\p@\scalebox{\@tempa}{\usebox\pandoc@box}%
  \else\usebox{\pandoc@box}%
  \fi%
}
\def\fps@figure{htbp}
\makeatother
\NewDocumentCommand\citeproctext{}{}

\makeatletter
 \let\@cite@ofmt\@firstofone
 \def\@biblabel#1{}
 \def\@cite#1#2{{#1\if@tempswa , #2\fi}}
\makeatother
\newlength{\cslhangindent}
\newlength{\csllabelwidth}
\newenvironment{CSLReferences}[2] % #1 hanging-indent, #2 entry-spacing
 {\begin{list}{}{%
  \setlength{\itemindent}{0pt}
  \setlength{\leftmargin}{0pt}
  \setlength{\parsep}{0pt}
  \ifodd #1
   \setlength{\leftmargin}{\cslhangindent}
   \setlength{\itemindent}{-1\cslhangindent}
  \fi
  \setlength{\itemsep}{#2\baselineskip}}}
 {\end{list}}
\usepackage{calc}

\newcommand{\CSLLeftMargin}[1]{\parbox[t]{\csllabelwidth}{\strut#1\strut}}
\newcommand{\CSLRightInline}[1]{\parbox[t]{\linewidth - \csllabelwidth}{\strut#1\strut}}

\usepackage{dcolumn, float, fancyhdr, hanging}
\graphicspath{{figures/}}
\usepackage{bookmark}
\IfFileExists{xurl.sty}{\usepackage{xurl}}{} % add URL line breaks if available
\hypersetup{
  pdftitle={Regime shifts and transformations in social-ecological systems:},
  colorlinks=true,
  linkcolor={Maroon},
  filecolor={Maroon},
  citecolor={Blue},
  urlcolor={blue},
  pdfcreator={LaTeX via pandoc}}

\title{Regime shifts and transformations in social-ecological systems:}
\usepackage{etoolbox}
\makeatletter
\providecommand{\subtitle}[1]{% add subtitle to \maketitle
  \apptocmd{\@title}{\par {\large #1 \par}}{}{}
}
\makeatother
\subtitle{Advancing critical frontiers for safe and just futures}
\author{\small Juan C. Rocha\(^{1,2,3,§}\), Caroline Schill\(^{1,2,4}\),
Emilie A. L. Lindkvist\(^{1}\), Reinette Biggs\(^{1,5}\),\\
\small Thorsten Blenckner\(^{1}\), Anne-Sophie Crépin\(^{1,4}\), Ingo
Fetzer\(^{1}\), Carl Folke\(^{1,2,4}\),\\
\small Amanda Jiménez-Aceituno\(^{1,6}\), Nielja Knecht\(^{1, 3}\), Jan
J. Kuiper\(^{1}\), Steven J. Lade\(^{1,7}\),\\
\small Carla Lanyon-Garrido\(^{1,8}\), Romi Lotcheris\(^{1}\), Romina
Martin\(^{1}\), Vanessa Masterson\(^{1,9}\),\\
\small Petr Matous\(^{10}\), Michele-Lee Moore\(^{1,11}\), Magnus
Nyström\(^{1}\), Per Olsson\(^{1}\),\\
\small Laura M. Pereira\(^{1,12}\), Garry Peterson\(^{1}\), André Pinto
da Silva\(^{1,3,13}\), Sasha Quahe\(^{1}\),\\
\small Maja Schlüter\(^{1}\), Lan Wang-Erlandsson\(^{1,2,3,13}\), and
Hannah Zoller\(^{1}\)\\
\strut \\
\tiny \(^{1}\)Stockholm Resilience Centre, Stockholm University, 10691
Stockholm, Sweden\\
\tiny \(^{2}\)Anthropocene Laboratory, The Royal Swedish Academy of
Sciences, 11418 Stockholm, Sweden\\
\tiny \(^{3}\)Bolin Centre for Climate Research, Stockholm University,
Sweden\\
\tiny \(^{4}\)Beijer Institute of Ecological Economics, The Royal
Swedish Academy of Sciences, 10405 Stockholm, Sweden\\
\tiny \(^{5}\)Centre for Sustainability Transitions (CST), Stellenbosch
University, South Africa\\
\tiny \(^{6}\)Social-Ecological Systems Institute (SESI), Leuphana
University of Lüneburg, Lüneburg, Germany\\
\tiny \(^{7}\)Fenner School of Environment \& Society, Australian
National University, Canberra, Australia\\
\tiny \(^{8}\)Energy Centre, Faculty of Physical and Mathematical
Science, Chile University, Santiago, Chile\\
\tiny \(^{9}\)Anthropology Department, Rhodes University, Makhanda,
South Africa\\
\tiny \(^{10}\)School of Project Management, University of Sydney,
Sydney, Australia\\
\tiny \(^{11}\)Dept of Geography and Centre for Global Studies,
University of Victoria, BC Canada\\
\tiny \(^{12}\)Global Change Institute, University of the Witwatersrand,
Johannesburg, South Africa\\
\tiny \(^{13}\)cE3c --- for Ecology, Evolution and Environmental Changes
\& CHANGE --- Global Change and Sustainability\\
\tiny Institute, Faculdade de Ciências, Universidade de Lisboa, Campo
Grande, Lisbon, Portugal\\
\tiny \(^{14}\)Potsdam Institute for Climate Impact Research, Member of
the Leibnitz Association, 14473 Potsdam, Germany\\
\tiny \(^{§}\)\href{mailto:juan.rocha@su.se}{\nolinkurl{juan.rocha@su.se}}}
\date{}

\begin{document}
\maketitle

\begin{abstract}
Current research challenges in sustainability science require us to consider nonlinear changes – shifts that do not happen gradually but can be sudden and difficult to predict. Central questions are therefore how we can prevent harmful shifts, promote desirable ones, and better anticipate both. The regime shifts and transformations literature is well-equipped to address these questions. Yet, even though both research streams stem from the same intellectual roots, they have developed along different paths, with limited exchange between the two, missing opportunities for cross-fertilisation. We here review the definitions and history of both research streams to disentangle common grounds and differences. We propose avenues for future research and highlight how stronger integration of both research streams could support the development of more powerful approaches to help us navigate toward safe and just futures.
\end{abstract}

\maketitle

\tableofcontents

\section{Introduction}

\begin{figure*}[!htb]
\centering
\includegraphics[width =0.95\textwidth]{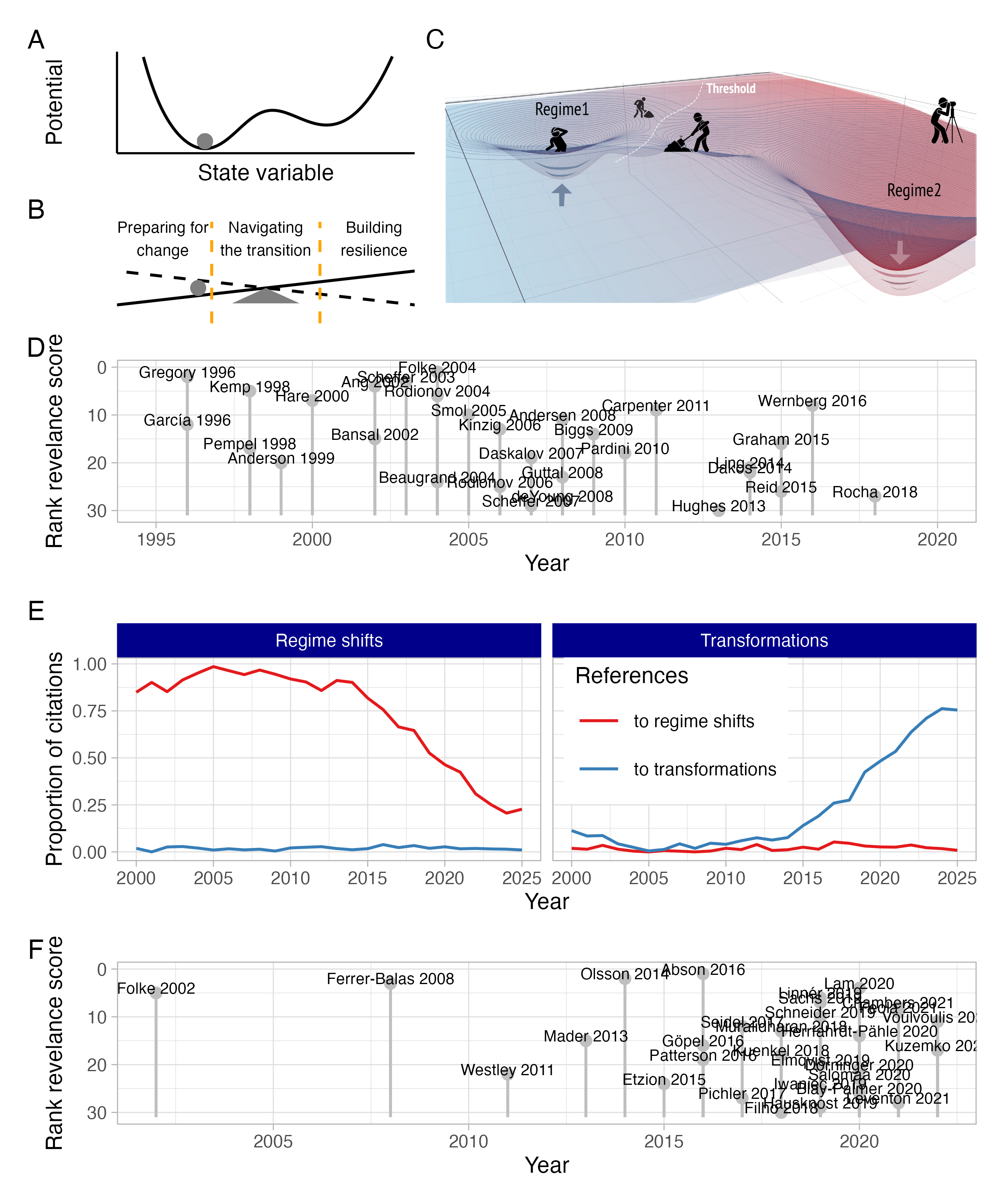}
\caption{\textbf{Regime shifts and transformation research.} The potential landscape has inspired heuristic models of regime shifts (A) and transformations (B), while regime shifts focus on plausible states of the system, transformations focus on imagining futures (C). The top 30 papers on regime shifts ranked by relevance are organized on a timeline for regime shifts (D) and transformations (F). Figure \ref{fig:top_papers} shows their bibliometric details. There is a lack of cross-fertilization between both strands of literature, with an annual proportion of references lower than 5\% and declining since 2020 (E), Figure \ref{fig:cross-ref} dissagregates the proportion per year. The idea of potential attractors is common to both, but there are differences on how they are studied. This represent an opportunity for cross-fertilization.} 
\label{fig:fig1}
\end{figure*}

Social-ecological systems research seeks to understand how social
actions influence ecological systems and, vice versa, how dynamics in
ecological systems affect social systems\textsuperscript{1}. These
systems are intertwined, and their mutual interactions are key for
sustained well-being of both ecosystems and people. Nonlinear changes
are of particular concern to sustainability scientists because they tend
to be harder to predict, cope with or adapt to, and costly and sometimes
impossible to reverse. They occur as a consequence of complex
social-ecological dynamics responding to disturbances in ways that are
not smooth, gradual, or proportional to the magnitude of the disturbance
(Fig \ref{fig:fig1}).

Two main strands of sustainability science address nonlinear changes in
social-ecological systems, namely the literature on regime
shifts\textsuperscript{2} and transformations\textsuperscript{3}.
Although the two concepts share a common origin\textsuperscript{4}, they
have diverged into different research strands over the past decades (Fig
\ref{fig:fig1}). While regime shifts are large, abrupt hard-to-reverse
changes in the function and structure of social-ecological
systems\textsuperscript{2,5}, transformations are often described as
involving intentional change that aim to fundamentally shift development
of a system into emergent pathways for particular goals, like
sustainability and justice\textsuperscript{6,7}.

The context in which both regime shifts and transformations take place
are social-ecological systems. Social-ecological systems (SES) are
intertwined, complex adaptive systems where people interact with
biophysical components. These systems often exhibit nonlinear dynamics
with threshold effects, feedback loops, time lags and heterogeneity in
system response and resilience\textsuperscript{1,8,9}. Resilience is the
ability of a system to deal with disturbances without permanently losing
its function, structure and identity\textsuperscript{4,6}. Research on
the resilience of social-ecological systems emphasizes three important
dynamic features: resilience as the capacity to cope with, adapt to, or
transform in response to change. Coping refers to the ability to absorb
or resist change to sustain current system dynamics; adaptability is the
capacity to learn and combine experience and knowledge to adjust
responses and system dynamics; while transformability is the ability to
cross thresholds and move the SES into emergent and often unknown
development trajectories\textsuperscript{6}.

The apparent separation between regime shifts and transformations
research motivates our review (Fig \ref{fig:fig1}, \ref{fig:cross-ref}).
Both regime shifts and transformations describe and are applied to
nonlinear changes between states of social-ecological systems and
emerged from research on resilience\textsuperscript{4}. However, these
research communities have developed independent conceptual frameworks,
methods and modes of application. This limited engagement risks missed
opportunities for cross-fertilization, reducing the potential to address
core challenges in sustainability science, which by definition requires
integration between social and biophysical dynamics. Similar reviews
have already considered the overlaps and distinctions between other
concepts, such as transitions and transformations\textsuperscript{10}.

In particular, dimensions that are central to one framework are at risk
of becoming oversimplified in the other. For example, agency ---the
capacity of actors to act upon their own choices --- is central within
transformations research\textsuperscript{3,11,12} but less prominent
within regime shifts studies, with some exceptions in
economics\textsuperscript{13}. Conversely, the role of feedbacks or time
scales are central in explaining regime shifts\textsuperscript{2,14,15},
but not always assessed in transformations research\textsuperscript{16},
with some exceptions\textsuperscript{17,18}. This lack of dialogue has
prevented the emergence of a consensus in terminology, leading to
oversimplification, and separate jargon denoting similar
characteristics. For example, the concept of positive tipping points
emphasizes desirable transformative changes such as the adoption of
electric vehicles\textsuperscript{19,20}, but often omits an analysis of
equity or which societal groups will benefit or suffer in this
transition, and what such changes implies for social-ecological
sustainability\textsuperscript{21,22}. In contrast, issues related to
justice and power have been studied for three decades in transformations
research\textsuperscript{21,23} highlighting that transformations always
involve shifting power relations and entail socially differentiated
risks and benefits; and can have unintended
consequences\textsuperscript{23}.

The parallel development of similar conceptual frameworks, without
scientific dialogue between them, risks reducing the capacity of
scientists to test their validity or adequacy in explaining real world
problems\textsuperscript{24,25}. As we will review, some of the
criticisms of regime shifts and transformations are lack of comparative
case studies that shed light under which conditions these frameworks
have explanatory power, or under which conditions are they likely to
fail. An additional risk of multiple untested frameworks is the use of
slightly different wording to mean the same thing (e.g.~positive tipping
points and transformations). Here, we will review and clarify the nuance
on meanings to avoid falling into the jargon trap, or at least make the
content more accessible from a pedagogical point of view. The next
section dives into the definitions and history of these concepts, trying
to disentangle what are their common grounds and differences. We then
outline current research frontiers and areas where we believe the two
strands of work could benefit from each other. Last, we conclude with a
short agenda for future research.

\section{History and definitions}\label{history-and-definitions}

The concept of resilience, as a fundamental characteristic of the regime
shifts and transformation frameworks, has its roots in mathematics. In
his seminal work introducing resilience, Holling\textsuperscript{4}
suggests minimal models to illustrate how ecosystems can change, lose
stability, and shift from one configuration to another, building on the
mathematical notion of bifurcations (e.g.\textsuperscript{26,27} ).
Several concepts such as the basin of attraction (or attractor), tipping
points, thresholds, or stability landscapes have precise definitions
within bifurcation theory, but are sometimes used as metaphors in later
conceptualizations. A basin of attraction, for example, is the region of
parameter space in which all trajectories converge to a common set of
points, the so-called attractor. Attractors can consist of single
equilibrium points, but they can also form more complex configurations,
such as (a)periodic cycles, long transients, or
chaos\textsuperscript{28}. We will not present formal definitions of
these concepts, but instead direct the interested reader to introductory
books on the matter\textsuperscript{28--30}.

In the mathematical context of dynamical systems, resilience has been
measured via the size of the basin of
attraction\textsuperscript{4,31--33}, the distance to its boundary, its
depth, or the rate of return to equilibrium after a disturbance, among
other measures\textsuperscript{32,34,35}. These definitions have enabled
the development of early warning signals for the proximity to a tipping
point based on fluctuations of the system state\textsuperscript{36}.
Bifurcation theory, and thus notions of resilience, have been
successfully applied to fields other than sustainability science and
ecology, including but not limited to language, climate systems, the
states of matter, cancer dynamics, finance, social dynamics such as
poverty traps, diffusion of beliefs, voting behavior, segregation, the
states of mind, or music, to name a few\textsuperscript{29,37}. Some of
these applications precede the work of Holling in
ecology\textsuperscript{4}, outlining the success of bifurcation theory
across the sciences. However, the evolutionary dynamics of adaptations
and transformations are not well defined within the seminal work on
bifurcations. While some formalizations have been used to explore for
example the evolution of innovations\textsuperscript{29,38,39}, it
remains an open problem to theorize true generative models that produce
agents with behaviors that were not previously coded by the
programmer\textsuperscript{40}. This problem is common ground to both
regime shifts and transformations research.

\subsection{Regime shifts}\label{regime-shifts}

Regime shifts are abrupt and persistent changes in the structure and
function of social-ecological systems\textsuperscript{2,5}. They occur
when resilience of the current regime is eroded and the system moves
from its original configuration into different modes of behavior.
Examples of regime shifts includes the transition from corals to
algae-dominated reefs\textsuperscript{41}, eutrophication of
lakes\textsuperscript{42}, the shift from forest to
savannahs\textsuperscript{43,44}, the disappearance of Arctic sea
ice\textsuperscript{45}, the collapse of the thermohaline
circulation\textsuperscript{46,47}, or the collapse of
cooperation\textsuperscript{48,49}. Regime shifts are important because
they often affect the flow of ecosystem services that people benefit
from\textsuperscript{50}, and they present a difficult management
problem since they can be hard to predict\textsuperscript{51,52} and
sometimes impossible to reverse on time scales relevant for management
and policy\textsuperscript{5,51}.

\subsubsection{Features}\label{features}

Regime shifts are typically characterized by feedbacks, tipping points,
thresholds, and hysteresis. Feedbacks are mutually interacting processes
that can amplify or dampen responses to drivers\textsuperscript{53}. For
example, moisture recycling feedbacks maintain tropical rain forests
like the Amazon: more forest transpires more humidity, feeding rainfall
that sustains the forest itself\textsuperscript{54}. In drylands,
facilitation feedbacks through shading, soil moisture and nutrient
retention are responsible for shifts between grasslands, shrublands and
deserts\textsuperscript{55,56}. In coral reefs, algae can preempt space
and increase sediment retention, reducing coral recruitment. As more
reefs shift to algal dominance, large-scale feedbacks such as regional
recruitment failure may reinforce local processes, reducing
fertilization success and preventing the regeneration of degraded coral
reefs\textsuperscript{14}. Or in social systems, poverty traps are
feedbacks mediated by income, education, or disease, that keeps people
in poverty despite development efforts\textsuperscript{57,58}. In
small-scale fisheries, poverty can lead to overfishing, which in turn
forces fishers to use more destructive fishing methods (e.g.~dynamite
fishing), locking the system in a vicious cycle\textsuperscript{59,60}.
While balancing feedbacks stabilize regimes, reinforcing feedbacks drive
out-of-equilibrium dynamics\textsuperscript{5,53}. Thus, weakening or
strengthening of feedbacks can trigger nonlinear transitions.

Tipping points are critical parameter values at which a system changes
its long-term qualitative behavior. They represent bifurcation points or
singularities\textsuperscript{27,28}. Speaking of a point implies the
assumption of a one-dimensional system with one co-dimension, that is a
very simplified version of reality where there is only one state
variable that we care about (e.g.~one species) and one driver affecting
it (e.g.~temperature). Real social-ecological system consist of multiple
interacting entities (e.g.~species, businesses, cities) that are driven
by multiple interacting forces, which are not necessarily reducible to
one (co)dimension. Complex bifurcations (of ``co-dimension'' greater
than one) can arise from the coincidence of multiple
drivers\textsuperscript{61,62}. When thinking of tipping phenomena in
real-life contexts, one should think of tipping lines or a hypersurface,
that include all the combination of critical parameters of all potential
drivers and state variables, rather than points. Stochastic variability
can further blur the ``point'' at which tipping happens. Tipping points
and thresholds are often used interchangeably. However, in mathematics
they have a slightly different meaning, the threshold is the separatrix
between two basins of attraction, while the tipping point is the
critical value at which a bifurcation occurs\textsuperscript{28} (Fig
\ref{fig:fig1}).

Observing and measuring tipping points or thresholds in real life is
challenging, leading some to view the concept not useful beyond
theoretical work\textsuperscript{63,64}. However, scientists have been
able to identify some tipping points with different degrees of
certainty. For example, hypoxia and anoxia are regime shifts common in
coastal areas where dissolved oxygen falls to levels that are not
tolerable for different groups of organisms. Levels below \(2mlO_{2}/L\)
gives rise to hypoxia, a condition where many fish die-off, levels below
\(0.5mlO_{2}/L\) trigger anoxia, a condition that is toxic also to
humans\textsuperscript{65,66}. Tipping points in precipitation have been
proposed and confirmed in observational studies for the transitions from
forest to savanna (\(1500mm/yr\)), savanna to forest (\(2000mm/yr\)),
and savanna to deserts (\(<500mm/yr\))\textsuperscript{43,44,67,68}.
Similarly, global thresholds for aridity has been reported for dryland
ecosystems (0.54, 0.7 and 0.8)\textsuperscript{69}, while thresholds for
deforestation in rain forest have been proposed (40\% for the Amazon,
25\% under precautionary principle, currently at
20\%)\textsuperscript{70,71}. Several Earth subsystems (e.g.~the
thermohalince circulation, the Greenland ice sheet) can have tipping
points driven by temperature. A recent literature review suggests that
six of these subsystems are highly likely to experience tipping dynamics
under \(2^\circ C\) of warming, with additional four of them tippping
under \(4^\circ C\) of warming\textsuperscript{72}.

Hysteresis is the reason why some regime shifts are difficult, sometimes
impossible to reverse (Fig \ref{fig:fig1}). Hysteresis means that
tipping point in one direction of the shift is different from the
tipping point in the opposite direction. Thus, additional efforts are
necessary to restore social-ecological systems after they have tipped,
and in some cases, full recovery may not be achievable at all. Much
research has focused on understanding, preventing and managing regime
shifts, and the interventions needed for recovery. Instead of listing
different strategies that apply to different ecosystems and contexts
(see box on the Regime Shifts Data Base), we summarize an approach
--adaptive management-- that has been successful in managing nonlinear
dynamics in the face of uncertainty.

Adaptive management was developed and advocated by early regime shift
scholars to encourage monitoring and experimentation, reduce
uncertainty, and inform decision making\textsuperscript{73,74}. It is
widely applied in fisheries, where monitoring and modeling are combined
to inform near-future maximum sustainable yields; projections that are
then updated through landing statistics and independent scientific
monitoring\textsuperscript{74}. Other applications are in fire
management, where prescribed burns prevent synchronized
fires\textsuperscript{75,76}; and in lakes, which can recover from
eutrophication after manipulating of nutrient, dredging, or
biomanipulation of the food web\textsuperscript{77--79}. Mathematical
models are often used to explore the space of management options, for
example to estimate critical nutrient loads in
lakes\textsuperscript{80,81}. In both cases, monitoring and
experimentation help managers get insights on relevant feedbacks or how
to adapt strategies and decisions to changing social-ecological systems.

\subsubsection{Historical developments and jargon
check}\label{historical-developments-and-jargon-check}

Inspired by bifurcation theory, the regime shift literature started off
with Holling's seminal paper on resilience and stability of
ecosystems\textsuperscript{4}. The key thesis was that ecosystems can
have multiple equilibria or steady states, and the probability of moving
from one to another was related to its resilience. It was a radical idea
at the time since most ecologists held the view that after disturbance,
nature would recover towards a pre-defined climax\textsuperscript{82}.
However, theoretical work that followed illustrated multiple stability
in communities\textsuperscript{83}, grazing systems\textsuperscript{84},
disease dynamics\textsuperscript{85}, fisheries\textsuperscript{86}, and
insect outbreaks\textsuperscript{87}. Empirical work in the following
decades provided evidence of regime shifts in
lakes\textsuperscript{77,79,88}, coral reefs\textsuperscript{89}, kelp
forests\textsuperscript{90,91}, grasslands\textsuperscript{92}, and
marine systems\textsuperscript{93,94}; and lead to the first syntheses
in the field\textsuperscript{2,95--99}.

These syntheses called for a more empirical grounding of the theory.
Even though there were some well documented case
studies\textsuperscript{77,79,88--94}, it was unclear to what extend
regime shifts were an exception or the rule\textsuperscript{97}. More
comparative case studies, experimentation and replication were
encouraged\textsuperscript{96}. It also became clear that if nonlinear
changes were a possibility in social-ecological systems, then we ought
to be designing management and policy strategies with that possibility
in mind. A body of work raised to the challenge and developed both
theoretical and empirical advances on how to manage regime
shifts\textsuperscript{2,50,100--105}. Key lessons included identifying
feedbacks\textsuperscript{104}, fast and slow processes that can be
influenced through interventions\textsuperscript{101}, and seizing
windows of opportunity when managerial actions have the most
impact\textsuperscript{106}. They also highlighted the value of
identifying and tracking key variables to anticipate thresholds and
inform management\textsuperscript{102,104}. Active participation by
communities and different actors was promoted, and adaptive
co-management methods and frameworks were developed to integrate
traditional ecological knowledge, indigenous and local practices into
knowledge generation and experimentation\textsuperscript{105}. These
bottom-up features are still practised and encouraged through
participatory resilience assessments\textsuperscript{107}.

The progress of bifurcation theory across the sciences, arts, and
humanities also resulted in multiple fields adopted slightly different
wording, or use the same wording to mean slightly different things.
These idiosyncrasies can be a source of confusion and regime shifts is
not the exception. In physics they are known as phase transitions, and
physicist distinguish first-order (discontinuous or with hysteresis)
versus second-order (continuous) transitions\textsuperscript{29}. Some
ecologist also call them critical transitions\textsuperscript{37},
although confusingly in physics the word critical is reserved to
continuous transitions (second-order), while ecologists typically use
the term to refer to bifurcations where hysteresis is plausible
(first-order). In coral reef science they are called phase
shifts\textsuperscript{108} and are expected to last at least at least 5
years\textsuperscript{109}, while in oceanography they are expected to
last decades and be driven by oscillations on climate systems (e.g.~the
Pacific Decadal Oscillation, the North Atlantic Oscillation, or El Niño
Southern Oscillation)\textsuperscript{110}.

Jargon and sometimes neglect of conceptual history have fueled
misconceptions. For example, in ecology, the concept of alternative
stable states seems to prompt people to think only of equilibrium
dynamics, and not of other out-of-equilibrium behavior as valid
regimes\textsuperscript{96,97,111}. However, both the mathematical
foundations and seminal regime shift papers consider cycles or chaos,
not only point equilibria, as regimes\textsuperscript{4,28,84--86}.
Likewise, hysteresis is sometimes oversimplified into too rigid
categories by conceptualizing regime shifts as either quasi-linear
change, reversible tipping (with no hysteresis, second-order
transitions), or irreversible tipping (with hysteresis,
first-order)\textsuperscript{29,95,98}. While this categorization is
useful from a pedagogical point of view, real world systems seldom
conform to these categories. A system that has only exhibited linear
changes in the past can exhibit hysteresis in the future under different
conditions (e.g.~new climatic conditions); and vice versa. For example,
simple models of disease spreading and (social) contagion dynamics
produce continuous outcomes, but models reflecting real world systems'
heterogeneity in the form of groups or networks give rise to
hysteresis\textsuperscript{112}. Similarly, simple models of transitions
in drylands and forest to savannas are discontinuous when analyzed in
one dimension, but they can become smooth when the dynamics are played
out in space\textsuperscript{113}. Despite hysteresis can emerge or
disapear in more complex systems, some climate scientist believe it to
be a necessary condition for tipping points\textsuperscript{20}.

Hysteresis is sufficient but not necessary for regime shifts, and theory
has shown, hysteresis can appear or disappear depending on context. In
practice, proving the existence of hysteresis implies running an
experiment to tip and recover a system, which is both impossible and
unethical in many real world situations. Models are insufficient
evidence of hysteresis since they only reflect the system dynamics
generated by a limited number of incorporated processes in a given
parameter space explored. As we will show in the next section,
hysteresis also depends on subjective system boundaries. However, even
when hysteresis has not been proven, continuous transitions are worth
considering as regime shifts as they can have long lasting impacts on
people, especially if reversibility occurs on time scales beyond human
lifespans or policy relevance.

\subsubsection{Empirical evidence and
detection}\label{empirical-evidence-and-detection}

Evidence of regime shifts has accrued over the years thanks to multiple
collective efforts. Most scientific articles document single case
studies or a set of cases of the same type of regime shift. This trend
called for a more systematic assessment and development of comparative
frameworks\textsuperscript{5}. Early efforts include the thresholds
database, which documents thresholds in social, ecological, and
social-ecological systems (N=103)\textsuperscript{114}. Other notable
community efforts are the global ocean oxygen network, which keeps track
of \textasciitilde400 know cases of coastal
hypoxia\textsuperscript{65,66}, the global field observations of forest
die-off (N=1303 plots)\textsuperscript{115}, the woody encroachment
database (N=499 places)\textsuperscript{116}, or the Chinese thresholds
database (N=110 cases)\textsuperscript{117}. To the best of our
knowledge, the Regime Shifts Database is the most comprehensive
repository of empirical evidence, synthesizing over 30 different types
of regime shifts using a common comparative framework, and over 3500
case studies around the world\textsuperscript{5}.

\fbox{\begin{minipage}{0.95\textwidth}
\subsection{The Regime Shifts Database}
Developed by the Stockholm Resilience Centre, the regime shifts database aims to synthesize current scientific knowledge on regime shifts in different social-ecological systems. The purpose of the database is to facilitate comparison and inform the general public on what is known on regime shifts, their potential impacts on society and management actions. It focuses on regime shifts that impact ecosystem services and human wellbeing. It only includes cases where clear feedback mechanisms have been proposed, and thus possibility for hysteresis exist, and that occur on a time scale relevant for managers and policy makers.

For generic regime shifts, the database offers a wikipedia style literature review of the main regimes and how the shift works, their drivers, impacts on ecosystem services and management options. It uses 75 categorical variables to enable comparison about their drivers, impacts, and key attributes such as the temporal and spatial scale of the shift, its reversibility, evidence type, confidence of the existence and mechanisms, as well as linkages to other regime shifts. The database uses causal loop diagrams to map the feedback structure of the system. A regime shift analysis provides a more technical explanation and uncertainty assessment of each feedback, driver and known tipping points. It further distinguish winners and losers of the shift and attempt to links impacts on sustainable devleopment goals.

Case studies are coded simiarly but with different levels of depth depending on the information available on scientific papers. Case studies are always linked to a scientific reference and when possible a precise time stamp and place where the regime shift has occurred. The database has been useful in international assessments of the Convention of Biological Diversity (CBD), the Intergovernmental Science-Policy Platform for Biodviersity and Ecosystem Services (IPBES), and the Arctic Resilience Assessment by the Arctic Council. Learn more at \href{www.regimeshifts.org}{www.regimeshifts.org}, or the development version at \href{regimeshifts.netlify.app}{regimeshifts.netlify.app} where new scientific and communicaiton features are tested.
\end{minipage}}

Despite the growing number of case studies and scientific studies,
evidence of regime shifts is still contested. This reflects both a
healthy skepticism about the theory's limits and a changing evidence
base and scientific consensus. For example, a decade ago, research on
West Antarctica's Ice sheet collapse provided conflicting arguments
about the existence and plausibility of tipping points. Today, there is
notably broader consensus, much more nuanced process understanding, and
even evidence that a tipping point may already have been
crossed\textsuperscript{118}. A counter example is summer Arctic sea ice
loss. Originally hypothesized as one of the classical tipping points in
climate\textsuperscript{45}, today questioned by some climate scientists
given the absence of hysteresis\textsuperscript{119}. Nonetheless, if
one changes the system's boundaries to include species depending on ice
ecosystems, or people whose livelihoods and culture depend on ice-scapes
(a social-ecological system), the shift is \emph{de facto} irreversible,
as it will cause extinctions or cultural loss beyond human generations.
Recent work on multi-lake datasets has also shown that many lakes do not
conform to the expectation of alternative stable
states\textsuperscript{120}. Yet, lack of evidence may stem from
insufficient data. In theory regime shifts are defined by the
relationship between independent or slow external drivers (such as
nutrient loading\textsuperscript{96}), but data is often only available
for integrative state variables (like nutrient concentrations). This
mismatch implies that available data may only provide a partial test of
the existence of shifting regimes.

Several efforts have been done in developing methods for detection and
prediction of regime shifts\textsuperscript{36,98,121--123}. Detection
methods include multivariate ordinations combined with different break
point analyses or bimodality tests\textsuperscript{98}. Log-response
ratios have been used in meta-analyses studying the effect of
environmental drivers on species, failing to find evidence of thresholds
due to poor signal to noise ratio\textsuperscript{63}. Increasing
temporal and spatial auto-correlation, and variance are common early
warning indicators of critical slowing down, a phenomenon describing the
decreasing recovery speed of complex systems in the run-up to regime
shifts\textsuperscript{36,121,122}. Other potential early warnings are
based on the analysis of, for example, skewness, flickering, kurtosis,
or specific patterns in Fisher information, fractal dimension (Hurst
exponents), patch size distributions, among others\textsuperscript{123}.
On the research challenges section we review current applications of
these methods and their success in predicting regime shifts.

\subsection{Transformations}\label{transformations}

Transformations are a fundamental restructuring of a system set of
relationships that hold a system in a particular
state\textsuperscript{124} or development pathway\textsuperscript{125}.
Transformations are multi-level and multi-phase processes that involve
changing the practices that influence the flow and distribution of
power, authority and resources\textsuperscript{124}; the norms, values
and beliefs underlying those practices\textsuperscript{11,126,127}, and
the way these connect to ecological systems across
scales\textsuperscript{128}. Examples of transformations includes the
French revolution, gaining of women voting rights, the end of slavery,
the end of Apartheid as transition towards democracy, or the shift from
command and control management towards community based governance in
Chilean fisheries\textsuperscript{9,99,129}. Transformations are
important because they are seen as a potential avenue to achieve
sustainable and just futures. In fact, major global assessment on
biodiversity and ecosystem services (IPBES) calls for transformative
change as and urgent an necessary way to avoid biodiversity loss and
potential collapse of ecosystem functions\textsuperscript{130}. IPBES
defines transformations as fundamental system-wide shifts in views,
structures and practices\textsuperscript{7}.

\subsubsection{Features}\label{features-1}

Ideas of attractors and crossing thresholds (Fig \ref{fig:fig1}) helped
conceptualize transformations in three phases: preparing for change,
navigating the transition, and building resilience of the new
regime\textsuperscript{3,131}. Feedbacks were important in the original
conceptualization to explore the role of agency and transformative
capacities. During the preparation-phase, unmaking self-reinforcing
feedbacks and weakening the structures that maintained the original
regime is essential. In contrast, creating a new regime requires
establishing new attractors, self-reinforcing feedbacks and
structures\textsuperscript{3}. The preparation and stabilization phases
are linked by the navigation phase when the system is in limbo and in an
in-between state\textsuperscript{124}. These feedbacks reflected
underlying norms, practices, values and beliefs that supported one
regime or the other\textsuperscript{11}. Transformation narratives have
often assumed normative goals and benefits\textsuperscript{23}, and as
such, they run the risk of reinforcing injustices by overlooking who
decides which aspects of a system should change, who leads the change,
and who benefits from it\textsuperscript{132}. Thus, to seek
representation, legitimacy and justice, sustainability transformations
typically rely on distributed agency, shift in power relationships, and
co-production approaches that enhance representation, legitimacy and
justice.

Agency is the ability to act, having the freedom to mobilize resources
and capacity to respond to challenges, and by doing so influencing other
agents and the environment\textsuperscript{133}. Agency, however, is a
contested concept with multiple interpretations across disciplines. In
sociology, agency is the ability of humans to act independently. It is
in tension with structural features of society (e.g.~norms, values,
social class, religion) which can constrain agency. In some disciplines
agency is associated to intentions, goal seeking, and free will
(e.g.~psychology), while in other disciplines it can be unintended,
unconscious, and attributed to non-human entities (e.g.~actor-network
theory or Indigenous cosmologies), or human collectives
(e.g.~corporations, nations). In transformations research agency is
distributed, meaning that no single author alone can drive
transformative systemic change\textsuperscript{124}. Transformations are
therefore problems of collective action\textsuperscript{134}.

Power is the relational capacity to bring about compulsory compliance,
it involves authority and the capacity to coerce\textsuperscript{135},
hence the uneven distribution of influence among actors over governance
process and objectives\textsuperscript{136}. In the context of
collective action, power structures can be illustrated on a situation
where the decisions on one user or group of users A affect group B more
than the decisions of group B affect group A\textsuperscript{135}. Thus,
power is embodied on social networks and can emerge from different
social attributes such as social structures (leadership, company board),
cultural attributes (elders, parenthood systems), market attributes
(concentration), technology (military), or context such as being on
upper stream versus downstream on a flowing resource (e.g.~water use in
water sheads, or fishers harvesting a migratory fish). In the context of
collective action, power is often studied as different forms of
heterogeneity that confers advantage to certain actors over
others\textsuperscript{134}. Too much heterogeneity, at leas in economic
terms (inequality), can decrease cooperation\textsuperscript{134,137},
although it remains an open question whether actors in positions of
privilege and power can steer systems towards sustainable development
when it is on their best interest\textsuperscript{138,139}. In the
context of transformations, power struggles occur between the actors
trying to keep the system in its current state and the actors seeking to
move it to a new one\textsuperscript{140}, as well as regarding whose
voice is heard when preparing for and navigating
transformations\textsuperscript{23}.

Justice is a core value to sustainability science. Sustainability is
about fulfilling our needs without compromising the ability of others in
current or future generations to do so. Justice is about the expectation
that everyone should be treated fairly. Justice has been conceptualized
as inter-species (with other living beings), intra-generational (within
the current generation, between communities or countries), and
inter-generational (with future generations)\textsuperscript{141}.
Recognition justice requires power structures and norms that marginalize
individuals or groups to be addressed; while epistemic justice
recognizes multiple forms of knowledge, including Indigenous, local
communities, and marginalized minorities, in science and decision
making\textsuperscript{141}. Transformation research typically uses
co-production approaches to research (e.g.~participatory action
research) to increase legitimacy, enhance representation, prevent
conflict, and promote more just futures\textsuperscript{124}.Historical
development and jargon check

\subsubsection{Historical development and jargon
check}\label{historical-development-and-jargon-check}

Inspired by regime shifts research, transformations research started off
with a series of seminal papers in early
2000s\textsuperscript{3,16,103,131}. Transformations research emerged
from an appreciation that in some instances adaptation or the
persistence of systems was inadequate to achieve aspirational goals with
the understanding that current historical, political and socio-economic
inequalities require a fundamental restructuring to achieve more
equitable outcomes\textsuperscript{142}. Thus transformations are deeply
ethical interventions\textsuperscript{143}. Transformations toward
sustainable and just futures necessitate significant shifts in power
dynamics, resource allocation, societal norms, and their connections
with ecological systems across different scales\textsuperscript{128}.

Concepts such as opportunity contexts emphasized the timing of
interventions, while leadership was thought as a capacity of key
individuals to bridge social networks and help gain momentum for the new
development pathways\textsuperscript{3,131}. More recent
conceptualizations pay lesser attention to leadership as a feature of
key individuals, and instead focuses on distributed agency of human and
non-human entities\textsuperscript{124}. They acknowledge that many
different actors have a role to play in the different phases of the
transformations, as the set of key actors and their interactions changes
during the process\textsuperscript{144--146}. Agency in transformations
can help trigger change or stabilize new regimes across
scales\textsuperscript{99}, by institutionalizing innovation, releasing
resources for innovation, or stimulating emerging innovations and
partnerships\textsuperscript{144}.

Tipping points in transformation research are often seen through the
lens of crises\textsuperscript{3,12,124,129,147}. Previous research also
frames them as critical junctures\textsuperscript{16,148}. Earlier
research suggested that a crisis could be an opportunity for
transformation (e.g.~by destabilizing the original regime or triggering
the emergence of shadow networks), but the scholarship has become more
nuanced on the multiple roles crises can play\textsuperscript{3,126}.
Crises can lead to transformation in one context, but `lock in' an
existing system in another\textsuperscript{149}. In non-scientific
literature tipping points are sometimes confused with particular people
or events\textsuperscript{150}, a misuse of the concept that has been
criticized in the social sciences\textsuperscript{25,151}. Tipping
points have been used also to understand behavioral change both in the
field and online social media platforms\textsuperscript{152--155},
theorize poverty traps\textsuperscript{57,58},
segregation\textsuperscript{156,157}, political
polarization\textsuperscript{158,159}, technology
adoption\textsuperscript{155}, inequality, or the emergence and collapse
of cooperation\textsuperscript{49,160}, to mention a few social
phenomena. All of these tipping phenomena can be enablers or barriers of
transformative change, but transformations are never determined by an
unique tipping point\textsuperscript{124}.

Attractors in transformation research are defined by what it could
be\textsuperscript{124,161}. Thus, narratives, visions and imagination
play a central role on how to explore plausible
attractors\textsuperscript{128,143,162,163}. The assumption is that if
people cannot imagine it, the attractor will not come to be. Therefore,
recent research has focused on what are these plausible futures and how
narrative methods and story telling can unlock some of that imagination
in particular contexts that include the oceans\textsuperscript{164},
food systems\textsuperscript{163}, or governing
biodiversity\textsuperscript{7}, or shifting world
views\textsuperscript{165}. A large scale exploration of plausible
futures is the Seeds of Good Anthropocene project where individuals and
communities explore marginal bright spots that exist today and could
scale up and become mainstream in the future\textsuperscript{166} (more
in the next section).

Before we explore some of the empirical evidence of transformations, it
is worth revisiting conceptual frameworks that share common features
with transformation research. The socio-technical transitions
framework\textsuperscript{167,168} has a stronger emphasis on
innovations at multiple scales and how dynamics at different scales of
aggregation can enable or restrict transformations. It has been applied
to sustainability problems related to energy, housing, water sanitation,
or food production\textsuperscript{167}. The three horizons
framework\textsuperscript{169} is a method inspired on transformations
research to explore futures and practices. Horizon 1 (H1) is the
currently dominant regime about to fade away, horizon 2 is the turbulent
transition phase where innovations are created, some of them supporting
H1 while others could lead to horizon 3 or the final regime of the
transformation. The three horizons framework has been used to explore
futures at the level of companies, non-governmental organizations,
goverment and academic actors\textsuperscript{169}.

Positive tipping points is another recent framework to think about
transformations\textsuperscript{19,170}. The concept has been widely
used in climate sciences to mean conditions and strategies for fast
decarbonization, including but not limited to, the adoption of new
technologies such as electric vehicles, solar panels, the
electrification of grids, or changing diets. The literature on positive
tipping points have been criticized for over-emphasizing technocratic
solutions while ignoring core issues of justice and distributions that
transformation research have studied for at least three
decades\textsuperscript{21}. Who defines what is positive, and positive
for whom, are political questions often ignored. The problem is that the
solutions landscape risks to reinforce unjust and colonial practices of
exploitation of minorities or less empowered. It can lead to temporary
or localized shifts (e.g.~more electric cars) without fundamentally
altering the underlying system dynamics, values and norms that
reinforced the undesired regime in first place\textsuperscript{171}.

\subsubsection{Empirical evidence}\label{empirical-evidence}

There is a growing body of research documenting evidence of
transformations. The Biosphere reserve Kristianstad Vatterike in
Southern Sweden was one of those iconic cases that helped early scholars
conceptualize transformations. It was established in 2005 and through
stewardship, co-management schemes, mobilization of shadow networks, and
the success of conservation stories such as the eco-museum, the
community demonstrated how to tackle the challenges of combining
conservation and development\textsuperscript{172--174}. Another
successful example is the transformation of marine coastal resource
governance in the coast of Chile\textsuperscript{125}. Through
co-management experiments and collaborations between fishers and
scientists, the governance system implemented an adaptive co-management
strategy that enabled the recovery of several fish stocks that crashed
in the late 1980s and early 1990s\textsuperscript{125}. In Panama,
researchers have shown how indigenous practices can be sources of
adaptive and transformative capacities showcasing an example of
combining multiple types of knowledge systems\textsuperscript{175}. In
South Africa, Sweden, and Uruguay transformations labs have been used to
address issues of food insecurity\textsuperscript{176--179}. These cases
showcase the use of dialogues, participatory action research, or arts
based methods to explore plausible futures and ways to get there.

Most of the scientific work however is dedicated to specific case
studies, there is no systematic comparative framework that is
continuously documenting past transformations (sensu Olsson \emph{et
al.}\textsuperscript{124}). The Arctic Resilience
Assessment\textsuperscript{180} developed a comparative framework
inspired by the regime shifts database to assess resilience and
transformations (N = 25), but at the time of publication only a handful
of them were coded. One example is Húsavik (Iceland) where the community
transformed their livelihoods from fishing and whaling to ecotourism. A
follow up study used an adaptive capacity framework\textsuperscript{9}
to code features that would increase or not the possibility of
transformations, finding that a strong sense of self-determination, the
ability to experiment and reuse previous knowledge and knowhow in new
alternative livelihoods were key in successful transformations (N=40
cases analyzed, 7 were transformations)\textsuperscript{181}. The recent
IPBES transformative change assessment identified around 400 local to
regional examples of transformative change, and compared the
environmental and social outcomes of these changes\textsuperscript{7}.
Many of these examples overlap with examples in the Seeds of Good
Anthropocenes dataset, which documents examples of potentially
transformative changes\textsuperscript{166}, not past transformations.

Transformation scholars not only study what has happened in the past,
but also the initiatives that today could become the mainstream of
future transformations\textsuperscript{182,183}. The Seeds of Good
Anthropocene\textsuperscript{166} is a collaborative project where
through participatory workshops scholars have been collecting local
ideas of sustainable initiatives that already exist but that are not yet
mainstream, but have transformative potential\textsuperscript{184--186}.
A recent analysis of African seeds showed that building support networks
is essential to foster transformative potential\textsuperscript{182}.
Futures-thinking is featured as a method that enables thinking in
potential plausible attractors\textsuperscript{187,188}, emphasizing
decolonial practices\textsuperscript{189,190}, and exploring the
dialogue with indigenous knowledge\textsuperscript{191}.

\fbox{\begin{minipage}{0.95\textwidth}
\subsection{Seeds of Good Anthropocenes}

The Seeds of Good Anthropocenes (SoGA) project is an international collaboration that seeks to broaden how people imagine the future. The seeds project is based on imagining and creating desirable, sustainable and just futures. A key part of this project is collecting and analyzing “seeds” —existing, non-mainstream real-world initiatives that people believe could grow to become part of a desirable sustainable future. These initiatives range from community-based conservation and alternative food systems to new forms of governance, technology, art, and education.

The project has built a global, open database of more than 500 such seeds (available at \href{goodanthropocenes.net}{goodanthropocenes.net}), that have been collected for related research projects connected to the SoGA project. Each seed entry includes information on its origins, context, and contributions to social and ecological well-being. Although each database has a specific focus, all databases are coded in a consistent fashion to enable comparisons of innovations across regions and themes, and to explore how different kinds of change can take root and spread. 

These scenarios aim to help people think creatively about transformation and highlight the diversity of pathways toward good Anthropocenes. SoGA’s methods have informed processes such as the IPBES Nature Futures Framework and have been applied in leadership training, education, and strategy development. The structure and design of this website benefited from the lessons of the regime shift database, in particular for managing the tradeoffs between the ease of contributing a seed, cohesion and consistency of contributions.

\end{minipage}}

\subsection{Similarities and
differences}\label{similarities-and-differences}

The literature on regime shifts and transformations have a common
history, common conceptual grounding. Our thesis, however, is that over
time the two strands have diverged (Fig \ref{fig:fig1},
\ref{fig:cross-ref}), thereby missing opportunities for
cross-fertilization and the risk of reinventing the wheel. One possible
reason for this divide could be that transformation research has been
traditionally geared towards the social sciences by focusing on
qualitative and narrative methods; while regime shift research has a
stronger use of mathematical analysis, computational modelling, and
controlled experiments. This is missed opportunity because it is now
recognized that the barrier between the social and natural sciences is
artificial and often counterproductive to advance the necessary
knowledge to achieve sustainability\textsuperscript{99,192--195}.

Both regime shifts and transformations are emergent phenomena; their
dynamics fall under the umbrella of complex adaptive
systems\textsuperscript{61}. In both cases, there are multiple
interacting elements (people, species, groups) each with their own
agency and constraints affecting each other and their environment. Both
are multidimensional problems that cannot be controlled with a single
master variable, or by a single actor --except perhaps in very simple
mathematical models used for pedagogical purposes. In both, collective
action and cooperation are required to move intentionally from one
regime to another. In both cases, history matters; the plausible future
pathways are in a way constrained by path dependency (e.g.~hysteresis).
The building parts of future innovations are features that already exist
today and have existed in the past, they succeed or perish following the
principles of evolution -- which is neither deterministic nor
teleological\textsuperscript{38--40}.

One misconception is that transformations are positive while regime
shifts are negative shifts. It is not that simple. First, the normative
value of the shifts depends on the eye of the observer, what is positive
for a group can be detrimental to another. That is why talking about
``positive'' tipping point can be misleading, the researcher risks
making the assumption that their level of desirability is the only valid
value, which can be problematic if not unethical. Transformation
research has dealt with that difference through the lens of backlash and
the potential conflicts and violence that can arise during the process
of transformation\textsuperscript{124}. The regime shifts database have
made an effort to identify winners and losers through their impacts on
ecosystem services they benefit from. However, this is not always
possible through literature synthesis efforts alone\textsuperscript{5}.
Second, things can go in unexpected ways. Transformations can end up on
an undesired regimes (at least for the majority). An example is the
current democratic backsliding, by 2019 two thirds of the countries
undergoing political regime transformations were becoming less
democratic\textsuperscript{196,197}. Some regime shifts can bring
desirable features, for example change in fire regimes and shifting from
coniferous to deciduous forest in boreal areas might increase in the
long term the capacity of the forest to store carbon in soils, helping
us contain the consequences of climate change\textsuperscript{198}.

One key difference between regime shifts and transformations research is
how researchers engage with the notion of attractors (or basins of
attraction). While both research streams use this notion as part of
their conceptualizations, the transformation literature is much more
intentional in studying the current narratives and imagined futures that
make up the plausible realization. In ecology, less attention is put on
the imagined nature of the attractor. For example, in restoration
policies, rewilding experiments, or invasive species management, there
is a socially constructed idea of what ``wild'' or ``invasive'' means,
the imagined meanings (or the meanings excluded) shaping the types of
policies implemented\textsuperscript{199}. As we enter the Anthropocene,
perhaps we need to reimagine what novel regimes could exist, and how our
management programs can already nudge change into more desirable and
inclusive pathways.

Another key difference is the style or mode of inquiry. Regime shift
scholars might situate themselves as observers analyzing the system from
the outside, while transformation scholars situate themselves playing an
active role within the system, helping navigate and taking part of the
process. We believe these two approaches are complementary and stronger
together. Quantitative analysis helps reveal patterns and tipping points
in system dynamics, while qualitative inquiry captures the meanings,
values, agency and contexts that shape how change unfolds. Together,
they provide the integrated understanding needed to push the frontiers
in each research field to achieve safe and just futures.\\

\section{Advancing research
frontiers}\label{advancing-research-frontiers}

Future research would benefit first from bridging both streams of
literature. The imaginary division between the social and the natural is
less than 300 years old and might be limiting our ability to create
actionable knowledge that rise to the challenges of the
Anthropocene\textsuperscript{195}. We propose five research frontiers
that could be advanced through cross-fertilisation: observability and
detection of non-linear system change, multidimensionality, evolutionary
dynamics, management, and ethics navigating the space of positive
tipping points.

\subsection{Observability and
detection}\label{observability-and-detection}

Both regime shifts and transformation research would benefit from
stronger efforts on observability, detection, and empirical evidence on
causation. Previous work has raised the issue that empirical evidence of
both regime shifts and transformations remains
scant\textsuperscript{63,96,97,181}. A few databases have been developed
but these are biased to the global north limited in
scope\textsuperscript{5,166}. To the best of our knowledge there is no
data repository on observed transformations, only for potential seeds.
Transformation and regime shift research is also biased towards
reporting true positives (cases where the phenomenon occurred), with
little documentation of true negatives, which are necessary for
comparison, causal inference, and theory development
(e.g.~counterfactuals). A balanced sample of true positives and true
negatives is necessary to compare features and help us explain what
factors made a transformation or regime shift possible. Some of the
methods recently used with such comparability in mind include
qualitative comparative analysis\textsuperscript{181,182}, and machine
learning approaches such as random forests\textsuperscript{200} or deep
neural networks\textsuperscript{201}.

\begin{figure*}[!htb]
\centering
\includegraphics[width =0.95\textwidth]{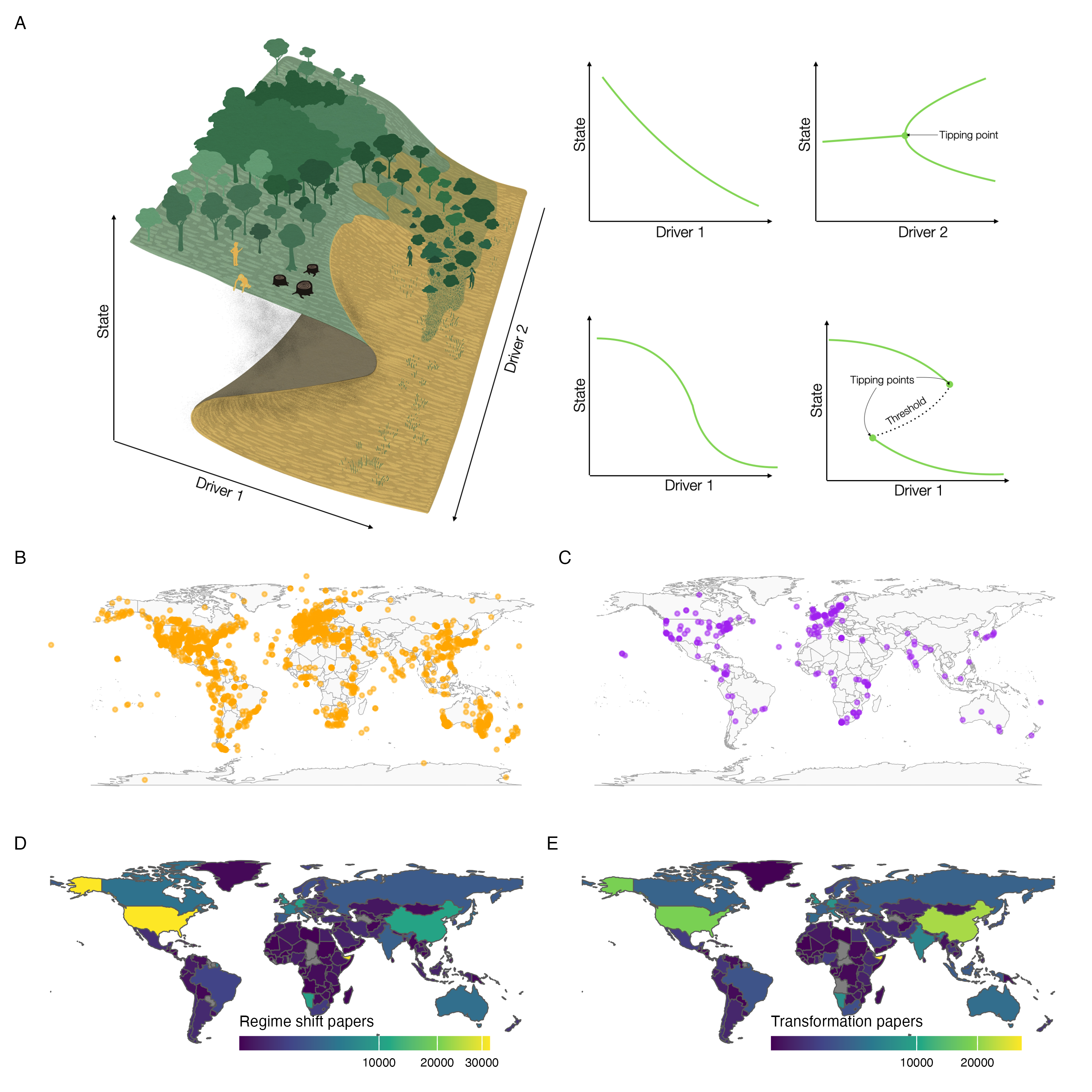}
\caption{\textbf{Empirical evidence} Both regime shifts and transformations used concepts from bifurcation theory as inspiration. A cusp manifold (A) shows different types of transitions using forest and savanna as example, depending on parameter values on one of the drivers (co-dim1) the transition can be smooth (top left), abrupt but without hysteresis (bottom left), or discontinuous with hysteresis (bottom right); but along the second driver (co-dim2) the transition is continuous (top right). Over 3500 cases of regime shifts are documented on the regime shifts database (B), while over 400 cases of potential transformations are documented on the seeds of good anthropocenes project which has informed the latest IPBES assessment (C). Academic production in number of academic papers is shown for regime shifts (D) and transformations (E) confirm the geographic bias in both areas.} 
\label{fig:fig2}
\end{figure*}

Beyond describing past events, a theory of regime shifts and
transformations should be capable of crafting expectations about the
conditions under which new realizations could occur. In other words,
there is a prediction problem. Regime shifts research has typically used
early warning signals\textsuperscript{36} to assess the probability or
proximity to tipping points, a proxy of resilience\textsuperscript{e.g.
68,200,202--205}. These papers have produced spatially explicit maps of
resilience that are not consistent across methods or data sets,
sometimes offering contradictory predictions or conclusions, and neither
of them making an effort to assess errors, accuracy or uncertainty. Most
papers rely on a single metric to draw conclusions, perhaps with the
exception of\textsuperscript{203} who use the synchrony between
autocorrelation and variance as proxy of quality of the signal. However,
recent work on high dimensions show that anti correlated signals can
emerge when multiple drivers or noise are involved in causing the
transition\textsuperscript{206,207}.

Assessing the efficacy of these methods on real world data remains a
scientific challenge. One needs a series of annotated data sets of true
positives, true negatives of the occurrence of events, and time series
data sampled at an adequate time scale to render useful signals. Recent
work shows that common early warning indicators perform poorly in
predicting forest die-offs events. Thus a key area of future research is
how to improve early warnings either by combining multiple signals, or
multiple methods, for example enhancing them with machine learning
approaches\textsuperscript{201}. One limitation of current machine
learning developments is that models are trained on synthetic data,
models that does not necessarily reflect the complexity of real world
settings.

Another open area of research is exploring how these early warning
signals could help identify windows of opportunity for transformations.
Transformation researchers are interested in identifying moments at
which an intervention such a policy or a social movement can amplify its
impact and move the system towards a different trajectory. If the
current system is too resilient, such efforts would be in vain. Thus,
early warnings signals could be used to identify windows of opportunity
where the seeds can have chance to challenge the status quo. For that
idea to work, transformation researchers will need to identify suitable
observables or variables that are informative of the dynamics of the
system. Some examples of potential variables include time series of key
commoditiy prices (e.g.~foods), voting participation, inequality, or
democratic sliding.

\subsection{Multidimensionality}\label{multidimensionality}

Complex adaptive systems are typically composed of multiple elements or
agents that interact at multiple temporal and spatial
scales\textsuperscript{61,99,208}. Yet most of the theory of regime
shifts and transformations relies on simplified assumptions involving
low dimensions and low co-dimensions (e.g.~only a few driving forces,
most frequently one). One outstanding challenge is how to navigate high
dimensionality in theory and practice\textsuperscript{209}. For example,
adequate models of ecological communities might easily have as many
dimensions as species in the food web, an informative model of a market
might require as many dimensions as products and services being
exchanged. There have been theoretical advances on dimension
reduction\textsuperscript{210,211}, but they are not always applicable
in real life settings.

One key question is what is the minimum number of dimensions that one
needs to still capture the key dynamics. If you were to observe a
social-ecological systems, what is the minimum number of observables one
needs to monitor to accurately gain insights on the dynamics of the full
system? These problems of observability of complex systems are not
new\textsuperscript{212}, but have not yet been explored in regime
shifts and transformation research. For example, designing early warning
signals for multivariate systems is still an open
question\textsuperscript{213}, with current approaches failing at
producing timely predictions\textsuperscript{214}.

Regime shifts can be interconnected, the occurrence of one regime shift
can impact the likelihood of another social-ecological system tipping
over\textsuperscript{15}. The mechanisms include domino effects when the
occurrence of one regime shift impacts the driver of another, or hidden
feedbacks when two regime shifts can amply or dampen each
other\textsuperscript{15}. These cascading effects have not been widely
studied in transformation research, but see\textsuperscript{215}. An
open area of research is how one transformation can enable or inhibit
other transformations, or create unjust settings in far away places.
Theoretical work has shown that coupled regime shifts leave no early
warning signals\textsuperscript{214}. In a high dimensional world, we
need to develop early warnings that are robust and still useful in
networked settings.

Transformation research has been concerned on how to engage with
plurality of knowledge\textsuperscript{216}, a perspective fully
embraced by the IPBES transformative change
assessment\textsuperscript{7}. Regime shifts research would benefit from
these methodological advances to deal better with social dynamcis
(e.g.~power, agency, justice) that might act as enablers or barriers to
ecosystem management and restoration. There is no right way to reduce
complexity, knowledge is always incomplete, and a diversity of
perspectives represent an opportunity to deal with social-ecological
challenges.

\subsection{Evolutionary dynamics}\label{evolutionary-dynamics}

Resilience scholars have adopted two key evolutionary features:
adaptability and transformability\textsuperscript{6}. However, the
ability to observe, measure and compare these features is rather
limited\textsuperscript{181}. As a result, most scholarly work treat
these features at the conceptual level\textsuperscript{99} with limited
ability to test or falsify theories. Both regime shifts and
transformation scholars would benefit from adopting a more rigorous
approach to study evolution in social-ecological systems. For example,
trait variation is essential in social-ecological feedbacks yet we know
little on how trait distributions impact tipping points, or how regime
shifts reshuffle trait dynamics\textsuperscript{217}. Theoretical work
has shown that fast evolutionary change with high trait variation can
reduce the sensitivity of ecosystems to rate-induced regime
shifts\textsuperscript{218}, or that evolution can delay the
manifestation of regime shifts\textsuperscript{219}. These early
examples show how including evolutionary process can change regime shift
dynamics, their path dependency, management strategies or their time
scales.

Principles of economic complexity and collective
learning\textsuperscript{220} can serve as an entry point to
operationalize evolutionary approaches in transformations research. For
example, if one think of transformations as shifting the livelihoods of
a community, there are datasets that inform the type of skills,
knowledge needed to gain competitive advantage on a series of
livelihoods (or producing commodities in a market), as well as the
transition probabilities from one set of livelihoods to another
(e.g.~labour statistics). Using this approach, researchers have make
inferences on the polarization of the work force\textsuperscript{221},
the resilience to financial shocks\textsuperscript{222},
inequality\textsuperscript{223}, or future economic
development\textsuperscript{224}.

\subsection{Management and governance}\label{management-and-governance}

Both regime shifts and transformations inherently poses the question of
who wins and who loses. Transformation research have shown that
conflicts of interest and power asymmetries between stakeholder groups
can give rise to conflict, push backs and even
violence\textsuperscript{124}. As such, one frontier of research is how
to develop capacities that help to navigate conflict and violence,
e.g.~peace building, conflict resolution, or methods to increase
inclusivity, legitimacy, and representation to navigate social backlash
dynamics\textsuperscript{124}. Other key capacities are hospicing or
honoring, grieving and addressing the losses of the dominant
system\textsuperscript{124,225}; or reflexivity to recognize the
constraints and opportunities of existing institutions in
place\textsuperscript{146}. How to develop and apply these capacities in
real world settings are open areas of research.

Interconnected regime shifts and transformations raises some additional
challenges for management and the capacity to navigate such
transitions\textsuperscript{226}. It means that optimal strategies in
one place might not be enough to trigger or avoid the transition
locally. A governance challenge of cooperation emerges: first one needs
to identify with whom does one need to cooperate to successfully manage
the SES; and second, it implies a cooperation problem with potential
power asymmetries\textsuperscript{226}. These power asymmetries might
hinder cooperation, exacerbate conflict or injustice. Preliminary
theoretical works shows that coupled interdependent regime shifts are
harder to manage that independent ones\textsuperscript{226}. Identifying
monitoring variables and empirically assess the coupling strength
between systems present opportunities for further
discovery\textsuperscript{15,226}.

Another interesting avenue for future research is how social-ecological
memory shapes management or option space. Previous research has shown
that the frequency of shocks such as fire determines the path dependency
of future forest dynamics creating ecological
memory\textsuperscript{75}. The lack of memory can prevent needed
transformation to avert unintended regime shifts, for example, the lack
of lifetime experience of Arctic variability or Amazon dieback. Most
regime shift research treats drivers as external forces to the system,
but sometimes shocks result from internal dynamics or feedbacks.
Studying how the frequency and intensity of shock disturbances affect
the management options of the system is then an open area of research.
For example, forest management can modulate water shocks (droughts and
floods) that may affect both transformative capacity and likelihood of
regime shifts. Another example is how the history of conflicts and
social traumas can hamper cooperation or the ability to imagine
alternative (e.g.~peaceful) futures.

\subsection{Ethics: Navigating positive tipping
points}\label{ethics-navigating-positive-tipping-points}

These management challenges do not come without ethical ones. Recently
some scientists have been advocating for the concept of positive tipping
points in the context of climate research and energy
transitions\textsuperscript{19,170}. While addressing climate change
with timely action is an imperative, simplified frameworks that ignore
the questions of positive for whom, or at the costs of whom, risk
advocating for superficial technocratic solutions that do not take
account of the required structural and values shifts for just and
sustainable transformations\textsuperscript{7,21,124,171}.

As such, another research challenge relates to the question of how to
navigate the transformations space (aka. positive tipping points) in an
ethical way\textsuperscript{21}. If the scientific enterprise is used to
advocate for quick fixes or to align with the interest of a few powerful
actors, it risks reinforcing colonial practices and power asymmetries
that marginalize minorities. It risks betraying the mantra of leaving no
one behind as inculcated in the UN 2030 Sustainable Development agenda.

\section{Conclusion}\label{conclusion}

Despite multiple historical attempts to encourage a scientific approach
that goes beyond the division of natural and social sciences, a truly
integrative approach is still lacking. We conclude that a truly
integrative approach is needed to rise to the challenges of the
Anthropocene. Integrating transformations and regime shifts research
might be a way forward. We outline a research agenda with common
challenges including improving the observation and detection techniques,
assessing multidimensionality, rigorously operationalizing evolutionary
dynamics, new management, governance, and ethical challenges.

Achieving sustainable development implies enabling just transformations
while staying within a safe operating space for humanity. It requires
skills and capacities to navigate regime shifts, avoid undesirable
tipping points when possible, implement timely adaptive management for
learning and understand what actions can make a difference.
Transformative change is likely to meet resistance, backlash, and even
violence. Integrating methods and techniques for conflict resolution,
increasing legitimacy, plurality of voices, and collective action is
likely to be key to enable safe and just futures.

\section*{Disclosure statement}

The authors are not aware of any affiliations, memberships, funding, or
financial holdings that might be perceived as affecting the objectivity
of this review.

\section*{Authorship statement}

All authors contributed to the conceptualization and writing review. JR,
EL, CS, GP, IF, AC, and M-LM contributed to methodology; JR, EL and CS
contributed with project administration, JR wrote the original draft,
developed software, analyzed data and created visualizations.

\section*{Acknowledgements}

The review emerged from a series of internal workshops at the Stockholm
Resilience Centre and a conference session at the Program for
Environmental Change and Society (PECS) in 2024. We would like to
acknowledge the contributions of workshops and conference session
participants including Beatrice Crona, Therese Lindahl, Liam
Carpenter-Urquhart, and Caroline Wallington. Eleanore Campbell supported
us with figure development. JCR was supported by grant 2022-04122 from
VR, the Swedish Research Council. SJL was supported by the Australian
Government (Australian Research Council Future Fellowship FT200100381).
JK and RM were supported by the PLURALAKES project, co-funded by the
Swedish Research Council FORMAS (no. 2024-00994) and EU's Horizon Europe
program through the 2023 Joint Transnational Call of the European
Partnership Water4All (no. 101060874). SQ was supported by the European
Research Council (ERC) under the European Union's Horizon Europe
research and innovation programme (grant agreement No 101097891 -
TRANSMOD). CS, EL, AC and JCR were supported by the Swedish government
research council for sustainable development (Formas), grant 2020-00454.

\section*{References}

\phantomsection\label{refs}

\newpage

\section{Supplementary Material}\label{sec:SM}

\renewcommand\thefigure{S\arabic{figure}}
\renewcommand\thetable{S\arabic{table}}
\setcounter{table}{0}
\setcounter{figure}{0}
\begin{figure*}[ht]
\centering
\includegraphics[width=0.95\textwidth]{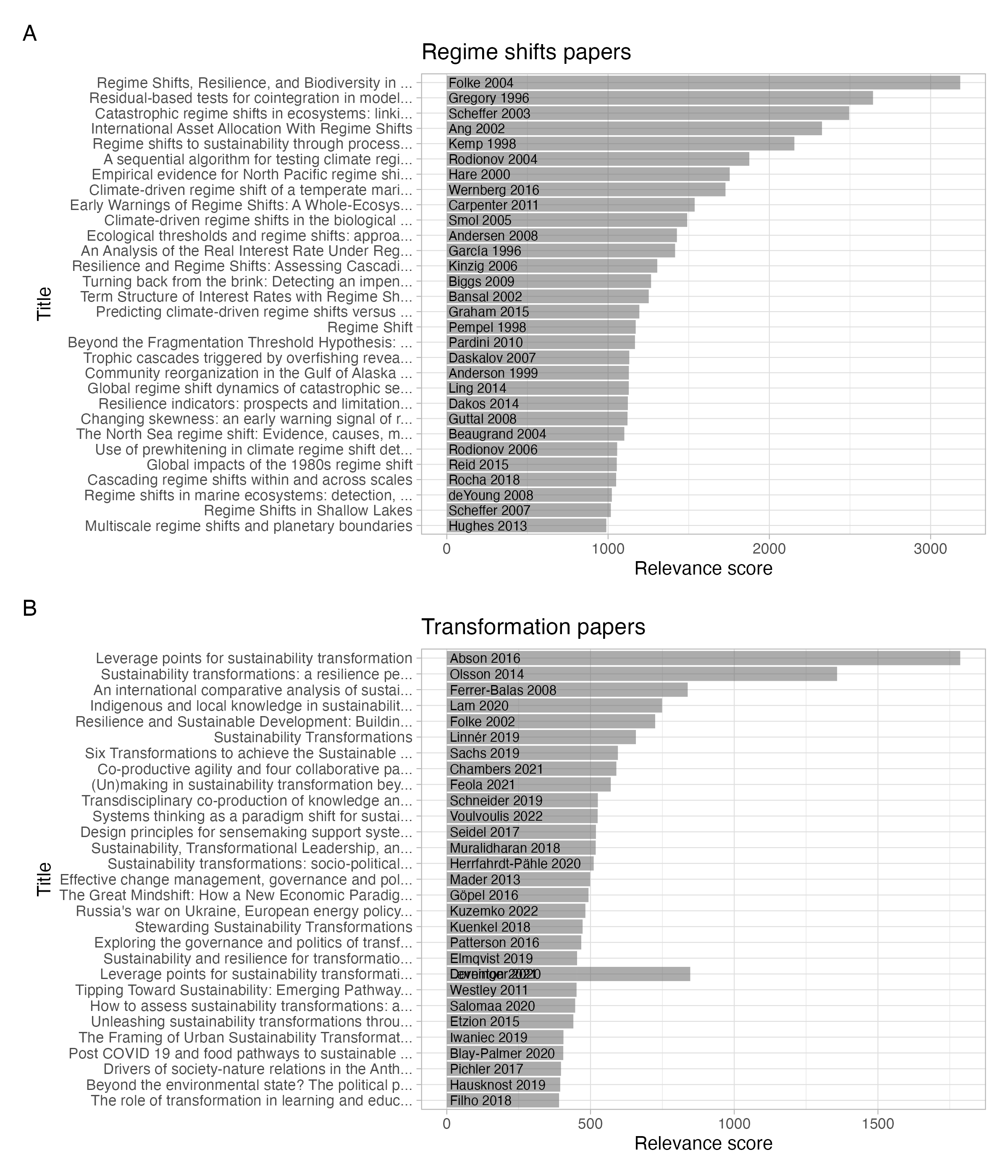}
\caption{\textbf{Top 30 papers} Data was sourced from OpenAlex with 53380 papers on regime shifts and 97300 papers on sustainability transformations, accessed in September 2025. OpenAlex is an open search engine for scientific literature. Relevance is computed as the semantic similarity between the search term (e.g. "sustainability transformations") and the text in the title and abstract, weighed by the number of citations.} 
\label{fig:top_papers}
\end{figure*}
\begin{figure*}[ht]
\centering
\includegraphics[width=0.95\textwidth]{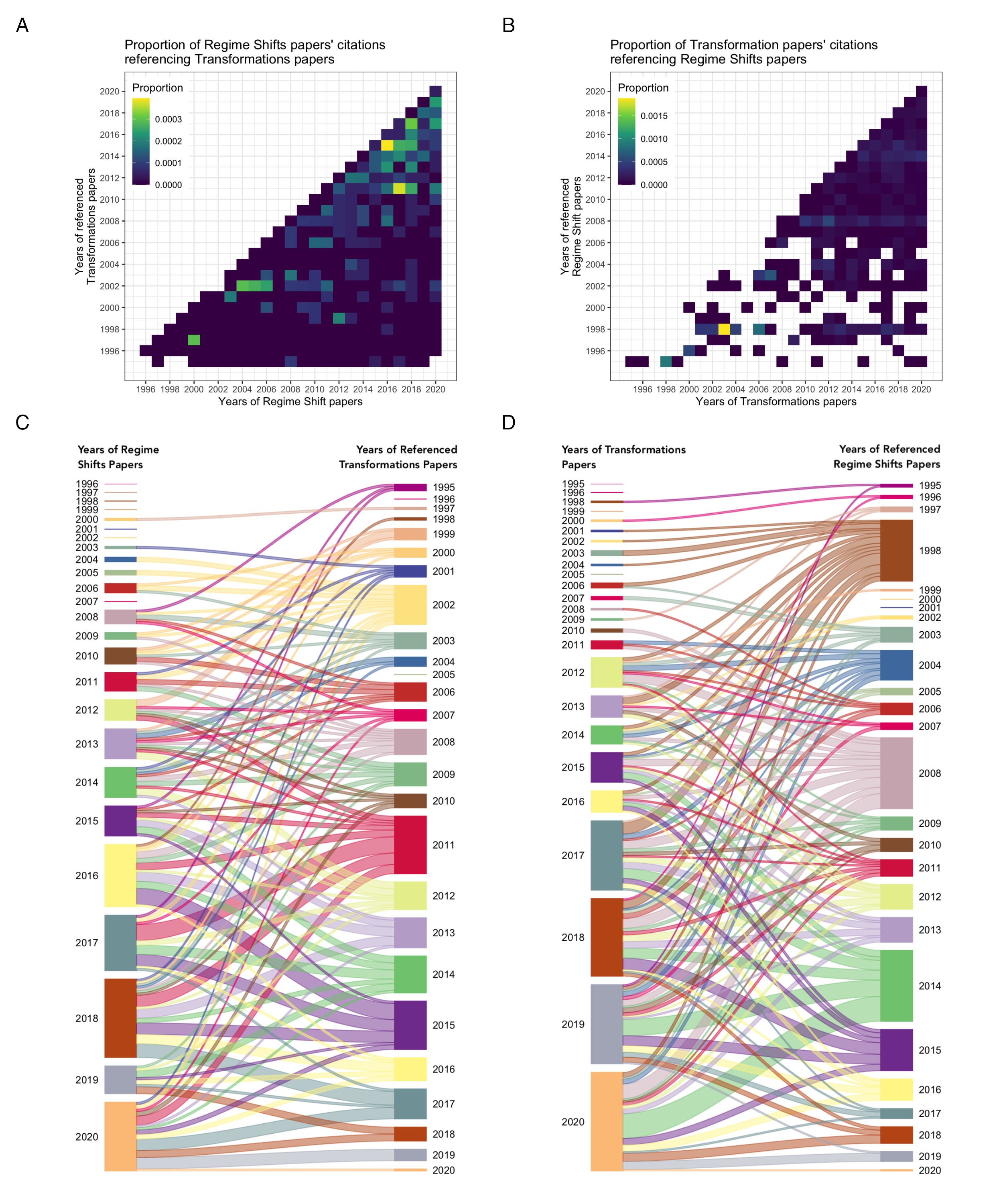}
\caption{\textbf{Lack of cross-fertilization} There has been low cross-referencing between the literature of regime shifts and transformations. Top panels present the proportion of references per year from one literature to another.} 
\label{fig:cross-ref}
\end{figure*}

\end{document}